\documentclass[submission,copyright]{eptcs}
\providecommand{\event}{PREPOST 2017} 
\usepackage{breakurl}             
\usepackage{underscore}           
\usepackage[pdftex]{graphicx}
\usepackage{listings}
\usepackage{color}

\definecolor{dkgreen}{rgb}{0,0.6,0}
\definecolor{gray}{rgb}{0.5,0.5,0.5} 
\definecolor{mauve}{rgb}{0.58,0,0.82}

\newenvironment{change}{}{}
  
\title{Fragmented Monitoring}
\author{Oscar Cornejo \qquad Daniela Briola \qquad Daniela Micucci \qquad Leonardo Mariani
\institute{Department of Informatics, Systems and Communication\\
University of Milan Bicocca 
Milan, Italy}
\email{\{oscar.cornejo, daniela.briola, daniela.micucci, leonardo.mariani\}@disco.unimib.it}
}
\def\titlerunning{Fragmented Monitoring}
\def\authorrunning{Cornejo, Briola, Micucci \& Mariani}
\begin{document}
\maketitle

\begin{abstract}

Field data is an invaluable source of information for testers and developers because it witnesses how software systems operate in real environments, capturing scenarios and configurations relevant to end-users. %
%
Unfortunately, collecting traces might be resource-consuming and can significantly affect the user experience, for instance causing annoying slowdowns.   

Existing monitoring techniques can control the overhead introduced in the applications by reducing the amount of collected data, for instance by collecting each event only with a given probability. However, collecting fewer events limits the amount of information extracted from the field and may fail in providing a comprehensive picture of the behavior of a program.


In this paper we present \emph{fragmented monitoring}, a monitoring technique that addresses the issue of collecting information from the field without annoying users. The key idea of fragmented monitoring is to reduce the overhead by recording partial traces (\emph{fragments}) instead of full traces, while annotating the beginning and the end of each fragment with state information. These annotations are exploited offline to derive traces that might be likely observed in the field and that could not be collected directly due to the overhead that would be introduced in a program. 


\end{abstract}

\section{Introduction}

Fully assessing the quality of software applications in-house is infeasible. Consider for instance the huge combination of configurations, inputs, and environments that should be verified. Since exhaustively testing every combination is not practical~\cite{Nie:Survey:CSUR:2011,Gazzola:ISSRE:2017,Gazzola:2017:FTS:3098344.3098487}, quality control activities cannot be limited to in-house verification and validation but must cross organization boundaries spanning the field.
This is why modern testing and analysis techniques often exploit data collected from the field, for example by collecting crash reports~\cite{Eclipse:website:2016,Windows:website:2016,Delgado:TaxonomyFaultMonitoring:TSE:2004} and profiling data~\cite{Elbaum:Profiling:TSE:2005}. 



While simple information could be retrieved quite efficiently without affecting the user experience, such as collecting a snapshot after a system crash~\cite{Eclipse:website:2016,Windows:website:2016,Delgado:TaxonomyFaultMonitoring:TSE:2004}, extensively collecting field data is challenging. %
In fact, monitoring executions, recording runtime data, and shipping the recorded information to developers are activities that interfere with regular executions. Indeed, they might introduce an annoying overhead and cause unacceptable slowdowns. Since preventing any interference with the user activity is a mandatory requirement, monitoring techniques address the challenge of collecting field data by reducing the amount of collected data in various ways. For instance, instead of collecting every relevant event, monitoring techniques may collect each event only with a given probability~\cite{Jin:Sampling:SIGPLAN:2010,Liblit:BugIsolation:SIGPLAN:2003}, or may collect only a subset of the relevant data by distributing the monitoring load across many instances of the same application~\cite{Orso:GammaSystem:ISSTA:2002,Bowring:MonitoringDeployedSoftware:PASTE:2002}. Unfortunately, in all these cases, monitoring can only retrieve a subset of the relevant data that might be extracted from the field to capture the behavior of the software. 

Consider for instance the problem of monitoring how some libraries are used within an application, that is, determining the sequences of calls to library methods produced by the application. Collecting events probabilistically would retrieve incomplete execution traces, which might be relatively useful if the events are collected with low probability. Collecting subsets of events also produces partial information about the execution scenarios, which might hinder the applicability of many analysis techniques.

In this paper, we present our initial effort in the design of a monitoring solution that can derive comprehensive runtime data without affecting the quality of the user experience. 
As a monitoring task, we consider the case of applications whose processes can be monitored independently. The target applications can thus range from centralized desktop applications to distributed server applications (as long as the system-level global ordering of events does not need to be extracted). 

Our key idea consists of collecting \emph{fragments}, that is, partial traces with no internal gaps, annotated with state information. Although fragments include only partial information about a monitored execution, they are complete when restricted to their time interval (i.e., every monitored event produced by the application after the first event of the fragment and before the last event of the fragment also occurs in the fragment). These fragments can be simply obtained by activating and deactivating a monitor that records every relevant event. %
Since the monitor is used intermittently, its impact on the user experience is limited. 

A fragment starts and ends with information about the state of the system. The idea is that by monitoring many executions and collecting many fragments, it should be possible to reconstruct traces that can likely be observed in the field from the fragments. In particular, fragments can be recombined offline consistently with the state information, that is, if a fragment ends in a state that matches the starting state of another fragment, the two fragments can be concatenated into a single fragment. In this way traces that extensively capture the behavior of a monitored program can be recreated from fragments, achieving lightweight monitoring without sacrificing the comprehensiveness of the collected data.

We called \emph{fragmented monitoring} the strategy for collecting and combining fragments that we designed to monitor the usage of one or multiple interfaces in a program. Fragmented monitoring uses an original combination of pre-deployment and post-deployment techniques. \emph{Before an application is deployed}, fragmented monitoring uses symbolic execution~\cite{King:Symbolic:ACM:1976,Braione:Enhancing:FSE:2013} to automatically identify the relevant program states that determine if and how an execution can proceed with additional calls to any of the monitored interfaces. This information is encoded as a set of conditions that can be used to discriminate the states of the monitored program, that is, two states that produce the same evaluations for these conditions are likely to use the monitored interfaces in a same way. Viceversa, states that produce different evaluations for these conditions are likely to use the monitored interfaces differently. 

The conditions computed before the software is deployed are embedded into the monitor and used in the field to derive the state information that annotates the fragments. The collected fragments are retrieved and combined \emph{post-deployment} to obtain the likely full traces. Although the reconstructed traces are not guaranteed to be actual traces of the monitored program, this strategy may enable a range of testing and analysis solutions that can exploit imperfect field data. 

The paper is organized as follows. Section~\ref{sec:approach} overviews the fragmented monitoring approach. Sections~\ref{sec:predeployment} and~\ref{sec:postdeployment} describe the pre-deployment and post-deployment analysis respectively. Section~\ref{sec:challenges} presents the main challenges of our ongoing work. Section~\ref{sec:related} discusses related work. Section~\ref{sec:conclusions} provides final remarks.

\section{Fragmented Monitoring} \label{sec:approach}

Fragmented monitoring has been designed based on the results of a preliminary study that we conducted to investigate the magnitude of overhead that can be tolerated by end-users when using their applications~\cite{NIER2017}. In this study we exploited the classification of system response time proposed by Seow~\cite{seow} to compute if and to what extent a monitoring activity might interfere with the user experience. Seow's classification indicates the maximum response time of several classes of operations (ranging from clicking on menu items to submitting complex queries). We used this information to study if a monitor that collects function calls might interfere with the quality of the user experience. 

Interestingly, we discovered that the \emph{overhead is distributed unevenly across operations}: a same monitoring activity (for example, collecting function calls) might produce a significantly different overhead (from negligible to more than 300\%) depending on the kind of operation executed by the software. For example, simple GUI operations like navigating through menu items generate fewer function calls, and thus show a smaller overhead, compare to computing complex mathematical computations in spreadsheet applications.
Moreover, we discovered that a \emph{non-trivial overhead} (usually up to 80\%) \emph{is often tolerated} by end-users. 

Fragmented monitoring leverages these results. Since a significant overhead can often be tolerated by users, fragments of non-trivial length might be feasibly collected from the field, but at the same time, since the impact of the monitoring activity changes with the operation that is executed, monitoring must be cleverly limited to prevent the introduction of an excessive overhead for specific operations.

  \begin{figure}[!ht]
  	\centering
  		\includegraphics[width=6in]{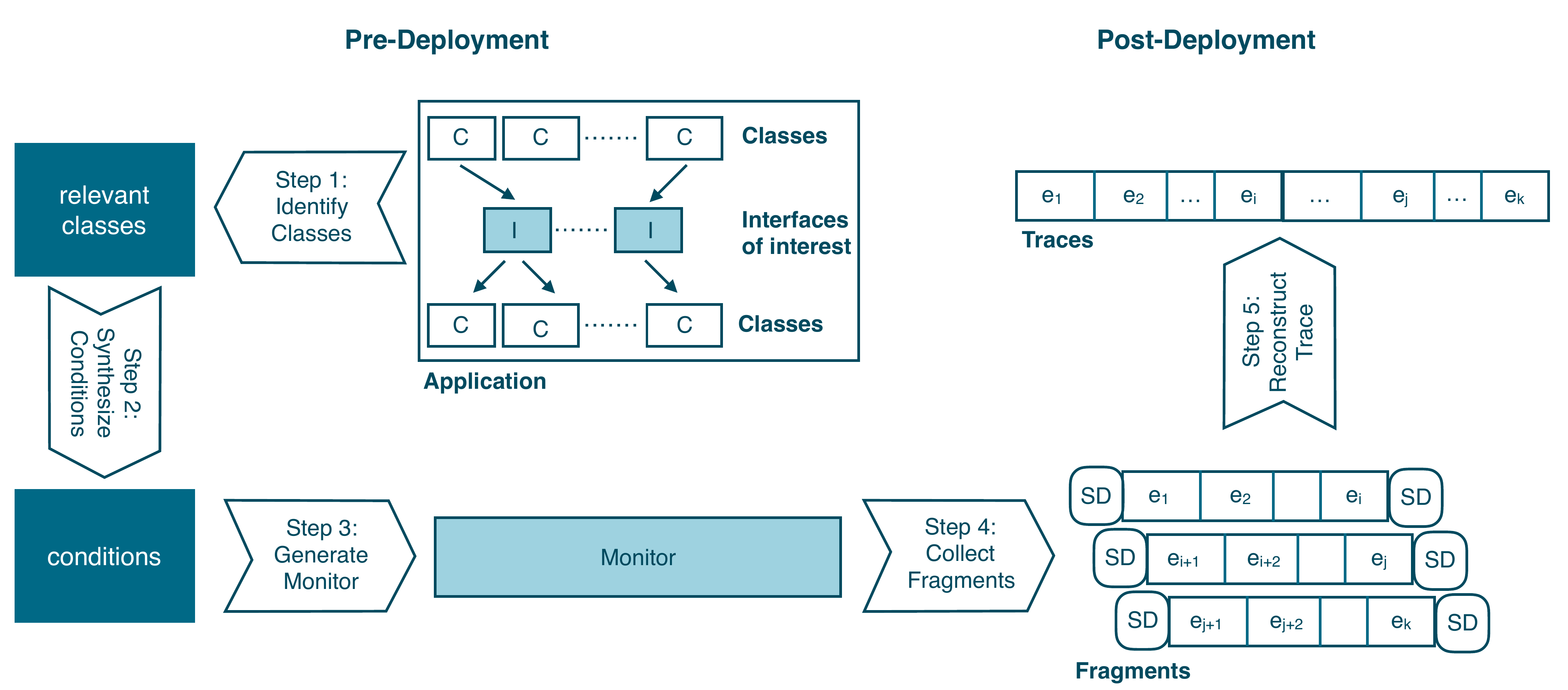}
  		\caption{Fragmented monitoring.}
  		\label{fig:approach}
 	\end{figure}

Figure~\ref{fig:approach} shows the overall fragmented monitoring process. Based on a set of \emph{interfaces of interest} (e.g., application APIs), fragmented monitoring first identifies the \emph{relevant classes} that must be monitored, that is, the classes that may influence the usage of the interfaces (step 1). In practice, the relevant classes are those that implement methods that can directly or indirectly invoke or be invoked by the methods defined in the monitored interfaces up to a user-customizable depth. Intuitively, these classes decide if and what interfaces methods to invoke based on the state of the application. 

In the second step, fragmented monitoring analyzes the relevant classes to distill the knowledge about the \emph{conditions} that may influence the execution, and thus influence the interaction with the monitored interfaces. For instance, let us assume that the monitored interfaces control the interaction with the file system: if a class using the monitored interface may actually open a file only if a string variable \texttt{text}, which represents the text to be written in the file, is non-empty, the condition synthesis step would output the condition \texttt{text.isEmpty()==false} to indicate that the evaluation of this condition may influence the interaction with the monitored interfaces.

All the conditions produced in step 2 are exploited to generate a Monitor (step 3) that collects fragmented executions. These steps are described in details in Section~\ref{sec:predeployment}.

Fragmented monitoring uses the generated conditions to produce a compact and relevant representation of the state of the program immediately before and after recording a fragment (step 4). We call this representation the \emph{signature data}. For instance, the condition \texttt{text.isEmpty()==false} would contribute to the signature data by producing a single value that depends on the length of the string. Note that these conditions generate good abstractions of the program state consistently with the conditions that might be evaluated by the program at any stage of the execution. In this case, since the program does not check the content of the string but only if it is empty, the signature data does not report information about its content. If the program would have processed the content of the string, 
the condition synthesis step would have also produced conditions for checking the content of the string consistently with the behavior of the program. 

The fragments are then analyzed offline. The signature data at the beginning and at the end of the fragment abstractly represent the state of the system at the time the fragment was recorded. Since a signature data directly depends on the conditions actually computed by the program while interacting with the monitored interfaces, the signature data is supposed to be representative of how the execution may proceed, specifically in terms of the interaction with the monitored interfaces. In principle, two fragments, one ending with a value of the signature data that is the same than the initial value of the signature data of the other fragment, could be sequentially composed to represent a longer execution that has been \emph{likely} produced by the monitored application (step 5). 

In the next sections, we describe more in details the pre-deployment and post-deployment analysis supported by an example.

\section{Pre-Deployment Analysis} \label{sec:predeployment}

This section describes the pre-deployment phase. The most important activity performed in this phase is the identification of the conditions that are later used to produce signature data, that is, the compact and abstract representation of the program state at the time the collection of a fragment is started or stopped.

We use a small example to present this case. Consider a Java program that includes the \texttt{Vehicle} class, which represents a generic vehicle considering its weight (attribute \texttt{weight}) and its maximum velocity (attribute \texttt{maxVel}), and a \texttt{VehicleService} class, which computes various statistics about a set of vehicles, such as their mean velocity and their mean weight. Let us assume that we want to collect data about how the program uses the \texttt{Vehicle} class, 
 focusing on the sequences of method calls produced by the program. \begin{change}In this case, the methods implemented by the \texttt{Vehicle} class represent the interface of interest.\end{change} Capturing every call to every method of the \texttt{Vehicle} class might be expensive for operations extensively using the public methods implemented in the \texttt{Vehicle} class. 

\begin{change}
\subsection{Step 1: Identify Classes}\label{subsec:step1}
\end{change}
Fragmented monitoring first 
identifies the relevant classes of the program, that is, the classes that directly or indirectly determine the sequence of invocations to the methods in the \begin{change}monitored interface.\end{change} 
These classes are determined statically in two stages. The first stage computes the call graph, which represents the calling relationships between the methods of the application. In our prototype we build the call graph using the WALA~\cite{WALA} static analysis tool. 

The second stage starts from methods in the monitored \begin{change}interface, \end{change}
in our example the public methods of the class \texttt{Vehicle}, and determines the callers and the callees of these methods, and the corresponding classes. The calling relationship is navigated transitively until a user-defined depth. In this example, if we bound the depth to $1$, both the \texttt{Vehicle} class, which is a callee at depth 1, and the \texttt{VehicleService} class, which is a caller at depth 1, would be selected.
Intuitively, the selected classes represent classes whose behavior is strongly correlated to the usage of the monitored classes. 

\subsection{Step 2: Synthesize Conditions}\label{subsec:step2}
Fragmented monitoring uses 
\begin{change}the relevant\end{change} classes to extract the conditions that represent how the values of the state variables may influence the execution of the monitored program. These conditions will be exploited at runtime to record state information efficiently, that is, representing the state of the application abstractly with a limited risk of losing information relevant to the program. 

For instance, if a program checks if the value of the weight of a vehicle is greater than 5000 to decide how to use that vehicle, it is enough to record whether the condition \texttt{weight > 5000} is true or not when saving state information, with no need of recording the actual weight, which would be irrelevant to determine how the execution may proceed. 

The conditions computed by the monitored program, and in particular the conditions computed in the relevant classes, are a set of conditions that can be naturally exploited to define an efficient abstraction strategy that can be applied at runtime to generate the signature data. To extract these conditions from code, we symbolically execute each method in the selected classed and use the path conditions resulting from the analysis of each method as conditions for state abstraction. Note that a path condition represents how the inputs to a method, that is, the values of the parameters and the values of state variables, relate to the execution of a specific path inside the method.

For instance, if the \texttt{VehicleService} class includes the following methods:

\begin{lstlisting}
	public void weight(){
		int weight = 0;
		
		weight += truck.getWeight();
		weight += van.getWeight();
		weight += car.getWeight();
		
		weightMean = weight / 3;
		
		if (weightMean > 5000)
			heavy = true;			
	}
	
	public void velocity(){
		int velocity = 0;
		
		velocity += truck.getMaxVel();
		velocity += van.getMaxVel();
		velocity += car.getMaxVel();
		
		velocityMean = velocity / 3;
		
		if (velocityMean > 110)
			fast = true;
	}
\end{lstlisting}

\noindent The resulting path conditions would include the following conditions:\\ 
 
\noindent \texttt{({VehicleService.truck.weight} + {VehicleService.van.weight} +} \newline\texttt{{VehicleService.car.weight}) / 3 > 5000}

\smallskip

\noindent \texttt{({VehicleService.truck.weight} + {VehicleService.van.weight} +} \newline\texttt{{VehicleService.car.weight}) / 3 <= 5000}
 
 \noindent \ldots 
 
\noindent \texttt{({VehicleService.truck.maxVel} + {VehicleService.van.maxVel} +} \newline \texttt{{VehicleService.van.maxVel}) / 3 <= 110}

 \noindent \ldots 

These conditions capture relevant state properties and represent the different execution paths, that in turn may produce different sequences of invocations to the monitored interfaces. 
Every time fragmented monitoring evaluates these conditions at runtime, a new instance of the signature data is produced.
 
We use the JBSE~\cite{Braione:Enhancing:FSE:2013} symbolic executor to compute the path conditions from the relevant classes, bounding the exploration of loops. The signature data that is incorporated before and after each fragment is obtained by evaluating at runtime the extracted conditions on the concrete state of the program.

\subsection{Step 3: Generate Monitor}\label{subsec:step3}

The conditions are used to generate a monitor that is able to collect fragments during program executions. The monitor will thus include the ability to start and stop the recording activity, in the example the ability to start and stop monitoring the calls to the \texttt{Vehicle} class, 
and will also include the capability to evaluate the conditions every time the recording is started or stopped. The activation and deactivation of the monitor might depend on multiple factors, such as the available resources and the nature of the operation that is executed. 

\section{Post-Deployment Analysis} \label{sec:postdeployment}

\begin{change}
This section describes the post-deployment phase. The most important activity performed in this phase is the reconstruction of traces from fragments. We use the same example of the Pre-Deployment analysis to present this phase.

\subsection{Step 4: Collect Fragments}\label{subsec:step4}
\end{change}

When the monitored program is executed, fragmented monitoring produces a number of partial traces consisting of an evaluation of the extracted conditions on the state of the program, a sequence of events, and again an evaluation of the conditions on the state of the program. Since we capture the state of the application by evaluating a set of conditions, 
we use a vector to store the evaluations at runtime. We call this vector the signature data, because it compactly represents the identity of the current state (e.g., like a signature does with people). 

Each condition is a three valued expression, since there are three possible outcomes: \texttt{true}, \texttt{false}, and \texttt{unknown}. A condition evaluates to \texttt{true} if the values of the program variables at the time the condition is evaluated satisfy the condition. Similarly, a condition evaluates to \texttt{false} if the values of the program variables at the time the condition is evaluated do not satisfy the condition. If some of the elements that appear in a condition cannot be evaluated, for instance because an object with an attribute that should be considered in the evaluation of a condition is null or even non existing, the condition evaluates to \texttt{unknown}. For example, the condition 

\smallskip
\noindent \texttt{({VehicleService.truck.weight} + {VehicleService.van.weight} +} \newline\texttt{{VehicleService.car.weight}) / 3 > 5000}
\smallskip

\noindent evaluates to \texttt{true} if the three weights are 6000, 5000 and 7000, while it evaluates to \texttt{false} if the three weights are 2000, 1000 and 5000. Finally, it evaluates to \texttt{unknown} if any of the \texttt{truck}, \texttt{van}, and \texttt{car} objects is \texttt{null}.

Each condition produces an evaluation that is stored in the signature data. 
Considering our example with the \texttt{Vehicle} class, a fragment of its execution could take the form

\smallskip
\noindent \begin{small}\texttt{(U, U, F, \ldots, T) car() car.setWeight() car.setVelocity() \ldots truck() (T, T, F, \ldots, T)}\end{small}
\smallskip



\noindent where the part between parentheses is the signature data, 
which includes the evaluation of every condition identified in the pre-deployment analysis (\texttt{T}, \texttt{F}, \texttt{U} stand for true, false, and unknown, respectively), and the sequence of labels between the two instances of the signature data is the sequence of events collected at runtime. 
If we assume that the condition whose evaluation produces the first element of the signature data is the aforementioned condition about the average weight of the three vehicles, the fragment indicates that the condition evaluated to unknown when the recording of the fragment has started, and evaluated to true when the recording of the fragment has been interrupted.

Fragmented monitoring accumulates fragments during the execution of the monitored application. Fragments can be collected both for multiple executions of a same application and for multiple executions of different installations of a same application. Collecting data from multiple installations of a same application simultaneously (e.g., multiple installations of a productivity suite) might increase by several orders of magnitude the amount of data available to reconstruct traces compared to the use an installation only.



%
%
%

\begin{change}
\subsection{Step 5: Reconstruct Traces}\label{subsec:step5}

This step 
\end{change}
processes the fragments recombining them in the attempt to obtain long and comprehensive traces. In this step, fragmented monitoring does not aim to obtain complete and fully sound traces, it rather aims at non-intrusively collecting useful traces from the field. 

To decide how to concatenate fragments, fragmented monitoring exploits the state information contained in the signature data, that is, if a fragment ends with signature data that matches the signature data reported at the beginning of another fragment, the two fragments are concatenated into a longer execution trace, which is still a fragment. Then, this process is iterated to obtain longer traces. 

\begin{change}
Suppose the following fragments have been collected:

\noindent \begin{small}\texttt{(U, U, F, \ldots, T) car() car.setWeight() car.setVelocity() \ldots truck() (T, T, F, \ldots, T)}\end{small}

\smallskip

\noindent \begin{small}\texttt{(T, T, F, \ldots, T) track.setWeight() track.setVelocity() \ldots (T, T, T, \ldots, T)}\end{small}

\smallskip

\noindent \begin{small}\texttt{(T, U, F, \ldots, T) van() van.setWeight() car() \ldots  (U, T, F, \ldots, T)}\end{small}
\smallskip

\end{change}
The first fragment could be concatenated with the second, but not with the third one (the signature data are different). The concatenation of the two fragments produces the following fragment:

\smallskip
\noindent \begin{small}\texttt{(U, U, F, \ldots, T) car() car.setWeight() car.setVelocity() \ldots truck() track.setWeight()} \newline \texttt{track.setVelocity() \ldots (T, T, T, \ldots, T)}\end{small}
\smallskip
 
In addition to exploiting the signature data (i.e., state information), fragmented monitoring can exploit additional sources of information to increase the precision of the reconstruction process. For example, it can exploit structural information, such as the control-flow graph of the program, to verify that a merged fragment still represents a feasible execution of the program.


\section{Challenges} \label{sec:challenges}

The design of fragmented monitoring is an ongoing effort. In order to transform this idea into a solid solution there are still several challenges that must be faced and that we discuss in this section.

\begin{itemize}
\item \textbf{\emph{Pre-deployment phase}} 

\begin{description}
\item[Complete condition synthesis:] Symbolic execution techniques can well serve the purpose of identifying the relevant conditions computed by a program, but the conditions resulting from this analysis may use method parameters in addition to state variables. While a condition that exclusively uses state variables might be simple to evaluate at any point of the execution, a condition that uses method parameters is not obvious to evaluate at an arbitrary point of the execution, as required by fragmented monitoring when producing the signature data. In fact a parameter is defined only when the corresponding method is invoked, while it does not exist in all the other points in time. Thus, a condition that includes a method parameter could be evaluated only if the monitoring is activated or deactivated with a call to the method that includes that parameter.

An obvious option is to drop conditions with method parameters, but this might cause the loss of too many conditions. Thus, the definition of strategies for rewriting conditions with parameters might be necessary to produce better sets of conditions.   

\item[Effective condition synthesis:] Not all the conditions are necessarily useful to abstract a concrete state, especially when monitoring focuses on some \begin{change}interfaces \end{change}
of interest only. Considering every condition returned by symbolic execution as a condition for producing the signature data might be too expensive, eliminating the benefit of producing fragments instead of full traces. The tradeoff between completeness and effectiveness must be carefully studied to define a strategy that identifies a small but effective set of conditions.  

\item[Monitor synthesis:] The generation of a monitor that evaluates the conditions at runtime is not a big challenge, but it might be challenging to define the strategy to decide when to start and stop fragments registration. If the fragments are too short, the cost of recording the signature data might be too high compared to the amount of information incorporated in a fragment. If the fragments are too long, users might experience some significant slowdowns. Finding strategies that well balance these two aspects is an open challenge. 

In addition to this, defining a strategy to start and stop the monitoring might be driven by the conditions that must be evaluated, and could consequently affect the effectiveness of the post-deployment phase. In fact, starting and interrupting the recording activities at some specific places could in principle make the post-deployment processing of the fragments simpler or harder.

\end{description}

\item \textbf{\emph{Post-deployment phase}}

\begin{description}
\item[Merging fragments:] We considered the signature data as the main information to concatenate fragments. However, there might be other strategies that can be exploited. For instance, fragments could be merged if they partially overlap in terms of the sequences of events included in the fragments and if the signature data satisfy some consistency rule weaker than equality. Even the concatenation may exploit a comparison criterion more flexible than equality. All these options require experimentation to be defined. 

\item[Exploiting additional information:] In principle, 
the post-deployment phase could exploit the source code of the program for distinguishing feasible and unfeasible traces produced by our approach. For instance, each trace could be checked against the control-flow of the program to verify if a specific combination of events could be feasibly produced. In general, static information could assist the process of reconstructing longer traces from fragments.  
\end{description}

\end{itemize} 
\section{Related Work} \label{sec:related}


\begin{change}

\end{change}

In the scope of techniques that exploit field data, Delgado et al.~\cite{Delgado:TaxonomyFaultMonitoring:TSE:2004} developed a taxonomy to analyze and differentiate runtime software \emph{fault-monitoring} approaches, by classifying elements considered essential for building a monitoring system. Likewise, Elbaum and Diep~\cite{Elbaum:Profiling:TSE:2005} assessed profiling techniques such as full, targeted and sampling profiling and their efficiency in testing activities.

Ohmann et al.~\cite{Ohmann:OptimizedCoverage:ASE:2016} worked on a technique for optimizing the placement of the instrumentation for collecting program coverage on applications running in the field. 
On the other hand, Jin and Orso~\cite{Jin:BugRedux:ICSE:2012} proposed BugRedux, an approach for supporting in-house debugging of field failures. In their work, they found that collecting method calls is the best type of data that can be extracted from the field to reproduce field failures in a controlled environment among the considered strategies. 

Clause and Orso~\cite{Clause:FieldFailuresDebugging:ICSE:2007} studied how to efficiently monitor the interactions between an application and its environment to determine the information that can be cost-effectively extracted from the field to reproduce failures. In the same perspective, Pavlopoulou and Young~\cite{Pavlopoulou:ResidualCoverage:ICSE:1999} presented a technique to validate in the field the functionalities that have not been fully tested in-house, specifically considering code coverage analysis.

Orso et al.~\cite{Orso:Selective:ACM:2005} also worked on a technique for a selectively capturing and replaying program executions. This technique allows to select a subsystem of interest, capture the interactions of the applications with its subystem, and then replaying the recorded interactions in a controlled environment. 

All these approaches either collect partial information without considering the problem of collecting comprehensive information about the runtime behavior of a program, or they assume that the overhead induced by the monitoring framework does not interfere with users. However, initial studies in this direction show that the monitoring overhead distributes unevenly across operations for identical monitoring tasks, sometimes, significantly impacting the user experience~\cite{NIER2017,Cornejo:Flexible:ICSEC:2017}. Ignoring this challenge may result in monitoring techniques that perturbate the interaction with the system annoying end-users.

\smallskip

The cost-effectiveness of the monitor is the main objective of our work. Current approaches dealing with this aspect exploits two kinds of solutions.

The first class is based on Distributive Monitoring, which divides the monitoring tasks between multiple installations of the same application, lowering the overhead introduced in each installation. 
This approach has been mainly developed by Orso et al.~\cite{Orso:GammaSystem:ISSTA:2002,Bowring:MonitoringDeployedSoftware:PASTE:2002} who defined the \emph{Gamma System} which relies on software tomography to optimize the placement of probes across several installations of the same application to minimize the runtime overhead. Bauer and Falcone~\cite{Bauer:Decentralised:FM:2012} considered a similar version of this problem by addressing the challenge of distributing the monitoring activities required for checking a formal specification in component-based systems, with the objective of avoiding to use a central data collection point.


The second class of solutions exploits Probabilistic Monitoring, which reduces the overhead by monitoring each instrumentation point only with a certain probability. Probabilistic monitoring thus reduces the overhead by collecting random subsets of the traces.
Liblit et al.~\cite{Liblit:BugIsolation:SIGPLAN:2003} have exploited probabilistic monitoring to isolate bugs by profiling a large, distributed user community and using logistic regression to find the important program predicates that could be faulty.
Similarly, Jin et al.~\cite{Jin:Sampling:SIGPLAN:2010} presented a monitoring framework called \emph{Cooperative Crug Isolation} to diagnose production run failures caused by concurrency bugs. This technique relies on sampling to monitor different types of predicates while keeping the overhead low.

A work similar to Probabilistic Monitoring, but in a different domain, is the one by Bartocci et al.~\cite{Bartocci:Adaptive:ICRV:2012}. They presented \emph{Adaptive Runtime Verification}, a monitoring technique that controls the overhead by enabling and disabling runtime verification of events according to overhead target levels. 
This framework determines the probability of an application property (based on observable actions of the monitored system) of being violated, and based on this value it assigns a higher or lower level of overhead to the monitored property. 
When verifying a certain property is not possible (e.g., due to a high overhead level) the framework estimates the probability that a property can be satisfied, instead of monitoring the actual property,  thus reducing the load of runtime verification.

Despite the benefits of these approaches, distributive and probabilistic monitoring have been developed with the goal of collecting partial information about the execution, while fragmented monitoring aims at deriving comprehensive traces.

\smallskip
Some monitoring tasks could be particularly expensive, for instance when the monitor collects parametric traces the runtime overhead introduced in the system might be significant~\cite{Chen:Parametric:TACAS:2009,Barringer:Quantified:FM:2012}. In this domain, the cost reduction that can be achieved with fragmented monitoring could be particularly relevant. In fact, systematically and extensively collecting parametric traces in object oriented applications can be hardly feasible without introducing sophisticated optimization strategies. 

In some cases, the optimizations could operate in the post-deployment phase rather than at runtime. For example, Diep et al.~\cite{Diep:Trace:ISSRE:2008} presented a technique for analyzing the traces produced by applications with the objective to identify and drop the redundant traces. As long as the relative frequency of traces is not concerned, this optimization strategy could be exploited in various contexts to reduce the cost of the post-deployment analysis, including fragmented monitoring.

\section{Conclusions} \label{sec:conclusions}

Collecting data from the field is challenging because monitoring solutions must satisfy two competing requirements: they must collect a relevant amount of data but they must also avoid interfering with the user activity. 
Since not interfering with users is usually the requirement with the highest priority, monitoring solutions tend to sacrifice the amount and kind of collected data in favour of a lower overhead. 

In this paper we presented our initial effort in the attempt to design \emph{Fragmented Monitoring}, an efficient monitoring solution that combines pre- and post-deployment analysis. The technique satisfies the need of collecting a limited amount of data per execution and the need of producing comprehensive information about the runtime behavior of an application. 
The key intuition underlying fragmented monitoring consists of recording executions fragments only, to limit the overhead, while annotating fragments with state information. The annotations enable the possibility to reconstruct the runtime behaviour of the system in a post-processing phase. 

The state information that must be sampled to annotate fragments is determined in a pre-deployment analysis phase that considers the structure of the program, together with the dependencies between the components and interfaces, as well as the conditions computed in the program.

Fully designing fragmented monitoring poses several challenges, some of which are still open. We plan to address these challenges in the near future and design experiments to compare the cost and effectiveness of fragmented monitoring to the cost and the effectiveness of other state of the art monitoring solutions. 

\section*{Acknowledgments}

This work has been partially supported by the H2020 Learn project, funded under the ERC Consolidator Grant 2014 program (ERC Grant Agreement n. 646867), and by the Italian Ministry of Education, University, and Research (MIUR) with the GAUSS PRIN project (grant n.~2015KWREMX).

\bibliographystyle{eptcs}
\bibliography{biblio}

\begin{thebibliography}{10}
\providecommand{\bibitemdeclare}[2]{}
\providecommand{\surnamestart}{}
\providecommand{\surnameend}{}
\providecommand{\urlprefix}{Available at }
\providecommand{\url}[1]{\texttt{#1}}
\providecommand{\href}[2]{\texttt{#2}}
\providecommand{\urlalt}[2]{\href{#1}{#2}}
\providecommand{\doi}[1]{doi:\urlalt{http://dx.doi.org/#1}{#1}}
\providecommand{\bibinfo}[2]{#2}

\bibitemdeclare{article}{Barringer:Quantified:FM:2012}
\bibitem{Barringer:Quantified:FM:2012}
\bibinfo{author}{H.~\surnamestart Barringer\surnameend},
  \bibinfo{author}{Y.~\surnamestart Falcone\surnameend},
  \bibinfo{author}{K.~\surnamestart Havelund\surnameend},
  \bibinfo{author}{G.~\surnamestart Reger\surnameend} \&
  \bibinfo{author}{D.~\surnamestart Rydeheard\surnameend}
  (\bibinfo{year}{2012}): \emph{\bibinfo{title}{Quantified event automata:
  Towards expressive and efficient runtime monitors}}.
\newblock {\sl \bibinfo{journal}{Proceedings of the International Symposium on
  Formal Methods (FM)}}, pp. \bibinfo{pages}{68--84},
  \doi{10.1007/978-3-642-32759-9_9}.

\bibitemdeclare{inproceedings}{Bartocci:Adaptive:ICRV:2012}
\bibitem{Bartocci:Adaptive:ICRV:2012}
\bibinfo{author}{E.~\surnamestart Bartocci\surnameend},
  \bibinfo{author}{R.~\surnamestart Grosu\surnameend},
  \bibinfo{author}{A.~\surnamestart Karmarkar\surnameend},
  \bibinfo{author}{S.~A. \surnamestart Smolka\surnameend},
  \bibinfo{author}{S.~D. \surnamestart Stoller\surnameend} \&
  \bibinfo{author}{J.n \surnamestart Seyster\surnameend}
  (\bibinfo{year}{2012}): \emph{\bibinfo{title}{Adaptive Runtime
  Verification}}.
\newblock In: {\sl \bibinfo{booktitle}{Proceedings of the International
  Conference on Runtime Verification (RV)}},
  \bibinfo{organization}{LNCS/Springer}, pp. \bibinfo{pages}{168--182},
  \doi{10.1007/978-3-642-35632-2_18}.

\bibitemdeclare{article}{Bauer:Decentralised:FM:2012}
\bibitem{Bauer:Decentralised:FM:2012}
\bibinfo{author}{A.~\surnamestart Bauer\surnameend} \&
  \bibinfo{author}{Y.~\surnamestart Falcone\surnameend} (\bibinfo{year}{2012}):
  \emph{\bibinfo{title}{Decentralised LTL monitoring}}.
\newblock {\sl \bibinfo{journal}{Proceedings of the International Symposium on
  Formal Methods (FM)}}, pp. \bibinfo{pages}{85--100},
  \doi{10.1007/s10703-016-0253-8}.

\bibitemdeclare{inproceedings}{Bowring:MonitoringDeployedSoftware:PASTE:2002}
\bibitem{Bowring:MonitoringDeployedSoftware:PASTE:2002}
\bibinfo{author}{J.~\surnamestart Bowring\surnameend},
  \bibinfo{author}{A.~\surnamestart Orso\surnameend} \& \bibinfo{author}{M.~J.
  \surnamestart Harrold\surnameend} (\bibinfo{year}{2002}):
  \emph{\bibinfo{title}{Monitoring deployed software using software
  tomography}}.
\newblock In: {\sl \bibinfo{booktitle}{Proceedings of the ACM SIGPLAN-SIGSOFT
  workshop on Program analysis for software tools and engineering (PASTE)}},
  \doi{10.1145/586094.586099}.

\bibitemdeclare{inproceedings}{Braione:Enhancing:FSE:2013}
\bibitem{Braione:Enhancing:FSE:2013}
\bibinfo{author}{P.~\surnamestart Braione\surnameend},
  \bibinfo{author}{G.~\surnamestart Denaro\surnameend} \&
  \bibinfo{author}{M.~\surnamestart Pezz{\`e}\surnameend}
  (\bibinfo{year}{2013}): \emph{\bibinfo{title}{Enhancing symbolic execution
  with built-in term rewriting and constrained lazy initialization}}.
\newblock In: {\sl \bibinfo{booktitle}{Proceedings of the Joint Meeting of the
  European Software Engineering Conference and the ACM SIGSOFT Symposium on the
  Foundations of Software Engineering (ESEC/FSE)}},
  \doi{10.1145/2491411.2491433}.

\bibitemdeclare{inproceedings}{Chen:Parametric:TACAS:2009}
\bibitem{Chen:Parametric:TACAS:2009}
\bibinfo{author}{F.~\surnamestart Chen\surnameend} \&
  \bibinfo{author}{G.~\surnamestart Rosu\surnameend} (\bibinfo{year}{2009}):
  \emph{\bibinfo{title}{Parametric Trace Slicing and Monitoring.}}
\newblock In: {\sl \bibinfo{booktitle}{Proceedings of the International
  Conference on Tools and Algorithms for the Construction and Analysis of
  Systems (TACAS)}}, \bibinfo{volume}{5505}, \bibinfo{publisher}{Springer}, pp.
  \bibinfo{pages}{246--261}, \doi{10.1007/978-3-642-00768-2_23}.

\bibitemdeclare{inproceedings}{Clause:FieldFailuresDebugging:ICSE:2007}
\bibitem{Clause:FieldFailuresDebugging:ICSE:2007}
\bibinfo{author}{J.~\surnamestart Clause\surnameend} \&
  \bibinfo{author}{A.~\surnamestart Orso\surnameend} (\bibinfo{year}{2007}):
  \emph{\bibinfo{title}{A technique for enabling and supporting debugging of
  field failures}}.
\newblock In: {\sl \bibinfo{booktitle}{Proceedings of the International
  Conference on Software Engineering (ICSE)}}, \doi{10.1109/ICSE.2007.10}.

\bibitemdeclare{inproceedings}{Cornejo:Flexible:ICSEC:2017}
\bibitem{Cornejo:Flexible:ICSEC:2017}
\bibinfo{author}{O.~\surnamestart Cornejo\surnameend} (\bibinfo{year}{2017}):
  \emph{\bibinfo{title}{Flexible in-the-field monitoring}}.
\newblock In: {\sl \bibinfo{booktitle}{Proceedings of the International
  Conference on Software Engineering (ICSE) - Companion}},
  \doi{10.1109/ICSE-C.2017.37}.

\bibitemdeclare{inproceedings}{NIER2017}
\bibitem{NIER2017}
\bibinfo{author}{O.~\surnamestart Cornejo\surnameend},
  \bibinfo{author}{D.~\surnamestart Briola\surnameend},
  \bibinfo{author}{D.~\surnamestart Micucci\surnameend} \&
  \bibinfo{author}{L.~\surnamestart Mariani\surnameend} (\bibinfo{year}{2017}):
  \emph{\bibinfo{title}{In the Field Monitoring of Interactive Applications}}.
\newblock In: {\sl \bibinfo{booktitle}{Proceedings of the International
  Conference on Software Engineering: New Ideas and Emerging Results Track
  (ICSE-NIER)}}, \doi{10.1109/ICSE-NIER.2017.19}.

\bibitemdeclare{article}{Delgado:TaxonomyFaultMonitoring:TSE:2004}
\bibitem{Delgado:TaxonomyFaultMonitoring:TSE:2004}
\bibinfo{author}{N.~\surnamestart Delgado\surnameend}, \bibinfo{author}{A.~Q.
  \surnamestart Gates\surnameend} \& \bibinfo{author}{S.~\surnamestart
  Roach\surnameend} (\bibinfo{year}{2004}): \emph{\bibinfo{title}{A taxonomy
  and catalog of runtime software-fault monitoring tools}}.
\newblock {\sl \bibinfo{journal}{IEEE Transactions on Software Engineering
  (TSE)}}, \doi{10.1109/TSE.2004.91}.

\bibitemdeclare{inproceedings}{Diep:Trace:ISSRE:2008}
\bibitem{Diep:Trace:ISSRE:2008}
\bibinfo{author}{M.~\surnamestart Diep\surnameend},
  \bibinfo{author}{S.~\surnamestart Elbaum\surnameend} \&
  \bibinfo{author}{M.~\surnamestart Dwyer\surnameend} (\bibinfo{year}{2008}):
  \emph{\bibinfo{title}{Trace normalization}}.
\newblock In: {\sl \bibinfo{booktitle}{Proceedings of the Symposium on Software
  Reliability Engineering (ISSRE)}}, \doi{10.1109/ISSRE.2008.37}.

\bibitemdeclare{misc}{WALA}
\bibitem{WALA}
\bibinfo{author}{J.~\surnamestart Dolby\surnameend}, \bibinfo{author}{S.~J
  \surnamestart Fink\surnameend} \& \bibinfo{author}{M.~\surnamestart
  Sridharan\surnameend} (\bibinfo{year}{visited on June 23 2017}):
  \emph{\bibinfo{title}{TJ Watson libraries for analysis (WALA)}}.
\newblock \bibinfo{howpublished}{\url{http://wala.sourceforge.net/}}.

\bibitemdeclare{misc}{Eclipse:website:2016}
\bibitem{Eclipse:website:2016}
\bibinfo{author}{\surnamestart {Eclipse Community}\surnameend}
  (\bibinfo{year}{visited on June 24 2017}): \emph{\bibinfo{title}{Eclipse}}.
\newblock \bibinfo{howpublished}{\url{http://www.eclipse.org}}.

\bibitemdeclare{article}{Elbaum:Profiling:TSE:2005}
\bibitem{Elbaum:Profiling:TSE:2005}
\bibinfo{author}{S.~\surnamestart Elbaum\surnameend} \&
  \bibinfo{author}{M.~\surnamestart Diep\surnameend} (\bibinfo{year}{2005}):
  \emph{\bibinfo{title}{Profiling deployed software: {Assessing} strategies and
  testing opportunities}}.
\newblock {\sl \bibinfo{journal}{IEEE Transactions on Software Engineering
  (TSE)}}, \doi{10.1109/TSE.2005.50}.

\bibitemdeclare{inproceedings}{Gazzola:2017:FTS:3098344.3098487}
\bibitem{Gazzola:2017:FTS:3098344.3098487}
\bibinfo{author}{L.~\surnamestart Gazzola\surnameend} (\bibinfo{year}{2017}):
  \emph{\bibinfo{title}{Field Testing of Software Applications}}.
\newblock In: {\sl \bibinfo{booktitle}{Proceedings of the International
  Conference on Software Engineering (ICSE) - Companion}},
  \doi{10.1109/ICSE-C.2017.30}.

\bibitemdeclare{inproceedings}{Gazzola:ISSRE:2017}
\bibitem{Gazzola:ISSRE:2017}
\bibinfo{author}{L.~\surnamestart Gazzola\surnameend},
  \bibinfo{author}{L.~\surnamestart Mariani\surnameend},
  \bibinfo{author}{F.~\surnamestart Pastore\surnameend} \&
  \bibinfo{author}{M.~\surnamestart Pezz{\`e}\surnameend}
  (\bibinfo{year}{2017}): \emph{\bibinfo{title}{An Exploratory Study of Field
  Failures}}.
\newblock In: {\sl \bibinfo{booktitle}{Proceedings of the International
  Symposium on Software Reliability Engineering (ISSRE)}}.

\bibitemdeclare{inproceedings}{Jin:Sampling:SIGPLAN:2010}
\bibitem{Jin:Sampling:SIGPLAN:2010}
\bibinfo{author}{G.~\surnamestart Jin\surnameend},
  \bibinfo{author}{A.~\surnamestart Thakur\surnameend},
  \bibinfo{author}{B.~\surnamestart Liblit\surnameend} \&
  \bibinfo{author}{S.~\surnamestart Lu\surnameend} (\bibinfo{year}{2010}):
  \emph{\bibinfo{title}{Instrumentation and Sampling Strategies for Cooperative
  Concurrency Bug Isolation}}.
\newblock In: {\sl \bibinfo{booktitle}{Proceedings of the ACM International
  Conference on Object Oriented Programming Systems Languages and Applications
  (OOPSLA)}}, \doi{10.1145/1869459.1869481}.

\bibitemdeclare{inproceedings}{Jin:BugRedux:ICSE:2012}
\bibitem{Jin:BugRedux:ICSE:2012}
\bibinfo{author}{W.~\surnamestart Jin\surnameend} \&
  \bibinfo{author}{A.~\surnamestart Orso\surnameend} (\bibinfo{year}{2012}):
  \emph{\bibinfo{title}{{BugRedux}: reproducing field failures for in-house
  debugging}}.
\newblock In: {\sl \bibinfo{booktitle}{Proceedings of the International
  Conference on Software Engineering (ICSE)}}, \doi{10.1109/ICSE.2012.6227168}.

\bibitemdeclare{article}{King:Symbolic:ACM:1976}
\bibitem{King:Symbolic:ACM:1976}
\bibinfo{author}{J.~C. \surnamestart King\surnameend} (\bibinfo{year}{1976}):
  \emph{\bibinfo{title}{Symbolic execution and program testing}}.
\newblock {\sl \bibinfo{journal}{Communications of the ACM}}
  \bibinfo{volume}{19}(\bibinfo{number}{7}), pp. \bibinfo{pages}{385--394},
  \doi{10.1145/360248.360252}.

\bibitemdeclare{article}{Liblit:BugIsolation:SIGPLAN:2003}
\bibitem{Liblit:BugIsolation:SIGPLAN:2003}
\bibinfo{author}{B.~\surnamestart Liblit\surnameend},
  \bibinfo{author}{A.~\surnamestart Aiken\surnameend}, \bibinfo{author}{A.~X.
  \surnamestart Zheng\surnameend} \& \bibinfo{author}{M.~I. \surnamestart
  Jordan\surnameend} (\bibinfo{year}{2003}): \emph{\bibinfo{title}{Bug
  isolation via remote program sampling}}.
\newblock {\sl \bibinfo{journal}{ACM SIGPLAN Notices}},
  \doi{10.1145/780822.781148}.

\bibitemdeclare{misc}{Windows:website:2016}
\bibitem{Windows:website:2016}
\bibinfo{author}{\surnamestart Microsoft\surnameend} (\bibinfo{year}{visited on
  June 24 2017}): \emph{\bibinfo{title}{Windows 10}}.
\newblock \bibinfo{howpublished}{\url{http://www.microsoft.com}}.

\bibitemdeclare{article}{Nie:Survey:CSUR:2011}
\bibitem{Nie:Survey:CSUR:2011}
\bibinfo{author}{C.~\surnamestart Nie\surnameend} \&
  \bibinfo{author}{H.~\surnamestart Leung\surnameend} (\bibinfo{year}{2011}):
  \emph{\bibinfo{title}{A survey of combinatorial testing}}.
\newblock {\sl \bibinfo{journal}{ACM Computing Surveys (CSUR)}}
  \bibinfo{volume}{43}(\bibinfo{number}{2}), pp. \bibinfo{pages}{11:1--11:29},
  \doi{10.1145/1883612.1883618}.

\bibitemdeclare{inproceedings}{Ohmann:OptimizedCoverage:ASE:2016}
\bibitem{Ohmann:OptimizedCoverage:ASE:2016}
\bibinfo{author}{P.~\surnamestart Ohmann\surnameend}, \bibinfo{author}{D.~B.
  \surnamestart Brown\surnameend}, \bibinfo{author}{N.~\surnamestart
  Neelakandan\surnameend}, \bibinfo{author}{J.~\surnamestart
  Linderoth\surnameend} \& \bibinfo{author}{B.~\surnamestart Liblit\surnameend}
  (\bibinfo{year}{2016}): \emph{\bibinfo{title}{Optimizing Customized Program
  Coverage}}.
\newblock In: {\sl \bibinfo{booktitle}{Proceedings of the IEEE / ACM
  International Conference on Automated Software Engineering (ASE)}},
  \doi{10.1145/2970276.2970351}.

\bibitemdeclare{inproceedings}{Orso:Selective:ACM:2005}
\bibitem{Orso:Selective:ACM:2005}
\bibinfo{author}{A.~\surnamestart Orso\surnameend} \&
  \bibinfo{author}{B.~\surnamestart Kennedy\surnameend} (\bibinfo{year}{2005}):
  \emph{\bibinfo{title}{Selective Capture and Replay of Program Executions}}.
\newblock In: {\sl \bibinfo{booktitle}{Proceedings of the International
  Workshop on Dynamic Analysis (WODA)}}, \doi{10.1145/1082983.1083251}.

\bibitemdeclare{inproceedings}{Orso:GammaSystem:ISSTA:2002}
\bibitem{Orso:GammaSystem:ISSTA:2002}
\bibinfo{author}{A.~\surnamestart Orso\surnameend},
  \bibinfo{author}{D.~\surnamestart Liang\surnameend}, \bibinfo{author}{M.~J.
  \surnamestart Harrold\surnameend} \& \bibinfo{author}{R.~\surnamestart
  Lipton\surnameend} (\bibinfo{year}{2002}): \emph{\bibinfo{title}{Gamma
  system: Continuous evolution of software after deployment}}.
\newblock In: {\sl \bibinfo{booktitle}{Proceedings of the ACM SIGSOFT
  international symposium on Software testing and analysis (ISSTA)}},
  \doi{10.1145/566172.566182}.

\bibitemdeclare{inproceedings}{Pavlopoulou:ResidualCoverage:ICSE:1999}
\bibitem{Pavlopoulou:ResidualCoverage:ICSE:1999}
\bibinfo{author}{C.~\surnamestart Pavlopoulou\surnameend} \&
  \bibinfo{author}{M.~\surnamestart Young\surnameend} (\bibinfo{year}{1999}):
  \emph{\bibinfo{title}{Residual test coverage monitoring}}.
\newblock In: {\sl \bibinfo{booktitle}{Proceedings of the International
  Conference on Software Engineering (ICSE)}}, pp. \bibinfo{pages}{277--284},
  \doi{10.1145/302405.302637}.

\bibitemdeclare{book}{seow}
\bibitem{seow}
\bibinfo{author}{S.~C. \surnamestart Seow\surnameend} (\bibinfo{year}{2008}):
  \emph{\bibinfo{title}{Designing and engineering time: the psychology of time
  perception in software}}.
\newblock \bibinfo{publisher}{Addison-Wesley Professional, ISBN: 0321509188,
  9780321509185}.

\end{thebibliography}
\end{document}